\documentclass[epj,nopacs]{svjour}
\usepackage{graphics}
\usepackage[english]{babel}

\begin{document}

\title{\boldmath Measurement of the $e^+e^- \to K^+ K^- \pi^0$  cross section
with the SND detector}

\authorrunning{M.~N.~Achasov et al.}
\titlerunning{Measurement of the $e^+e^- \to K^+ K^- \pi^0$  cross section}

\author{{\large The SND Collaboration}\\ \\
M.~N.~Achasov\inst{1,2} \and
A.~Yu.~Barnyakov\inst{1,2} \and
M.~Yu.~Barnyakov\inst{1} \and
A.~A.~Baykov\inst{1,2} \and
K.~I.~Beloborodov\inst{1,2} \and
A.~V.~Berdyugin\inst{1,2} \and
D.~E.~Berkaev\inst{1,2} \and
A.~G.~Bogdanchikov\inst{1}\and
A.~A.~Botov\inst{1}\and
A.~R.~Buzykaev\inst{1}\and
T.~V.~Dimova\inst{1,2} \and
V.~P.~Druzhinin\inst{1,2} \and
V.~B.~Golubev\inst{1} \and
L.~V.~Kardapoltsev\inst{1,2} \and
A.~G.~Kharlamov\inst{1,2} \and
A.~A.~Korol\inst{1,2} \and
D.~P.~Kovrizhin\inst{1} \and
E.~A.~Kravchenko\inst{1,2} \and
A.~S.~Kupich\inst{1}\and
A.~P.~Lysenko\inst{1}\and
K.~A.~Martin\inst{1}\and
N.~Yu.~Muchnoy\inst{1,2} \and
N.~A.~Melnikova\inst{1}
A.~E.~Obrazovsky\inst{1}\and
A.~P.~Onuchin\inst{1}\and
E.~V.~Pakhtusova\inst{1}
E.~A.~Perevedentsev\inst{1,2}\and
K.~V.~Pugachev\inst{1,2}\and
Y.~S.~Savchenko\inst{1,2}\and
S.~I.~Serednyakov\inst{1,2} \and
P.~Yu.~Shatunov\inst{1}\and
Yu.~M.~Shatunov\inst{1,2} \and
D.~A.~Shtol\inst{1} \and
D.~B.~Shwartz\inst{1,2} \and
Z.~K.~Silagadze\inst{1,2} \and
I.~K.~Surin\inst{1}\and
Yu.~A.~Tikhonov\inst{1,2}\and
Yu.~V.~Usov\inst{1} \and
I.~M.~Zemlyansky\inst{1,2}\and 
V.~N.~Zhabin\inst{1} \and
V.~V.~Zhulanov\inst{1,2}}

\institute{Budker Institute of Nuclear Physics, SB RAS, Novosibirsk, 630090, Russia
\and
Novosibirsk State University, Novosibirsk, 630090, Russia}
\date{}

\abstract{
The process $e^+e^-\to K^+K^-\pi^0$ is studied with the SND detector at the VEPP-2000 
$e^+e^-$ collider. Basing on data with an integrated luminosity of
26.4~pb$^{-1}$ we measure the $e^+e^-\to K^+K^-\pi^0$
cross section in the center-of-mass energy range from 1.28 up to 2 GeV.
The measured mass spectrum of the $K\pi$ system indicates that the dominant
mechanism of this reaction is the transition through the
$K^{\ast}(892)K$ intermediate state. The cross section for the
$\phi\pi^0$ intermediate state is measured separately. The SND results 
are consistent with previous measurements in the BABAR experiment
and have comparable accuracy. We study the
effect of the interference between the $\phi\pi^0$ and $K^\ast K$ amplitudes.
It is found that the interference gives sizable contribution
to the measured $e^+e^- \to \phi \pi^0\to K^+K^-\pi^0$ cross
section below 1.7 GeV. 
}

\maketitle

\section {Introduction}
This paper is devoted to the study of the reaction $e^+e^- \to K^+ K^- \pi^0$
in the experiment with the SND detector at the VEPP-2000 $e^+ e^-$
collider~\cite{bib:vepp2000}. This reaction is one of three charge
modes of the process $e^+ e^- \to K \bar{K} \pi $, which gives a
sizable contribution (about 12\% at the center-of-mass (c.m.) energy 
$\sqrt{s} \approx 1.65$ GeV) to the total cross section of $e^+ e^-$ 
annihilation into hadrons, and is the key process for measuring the 
$\phi(1680)$ resonance parameters. The reaction $e^+ e^- \to K^+ K^- \pi^0$
was first observed in the DM2 experiment~\cite{bib:dm2}. The accuracy of 
measuring its cross section was significantly improved in the BABAR 
experiment~\cite{bib:babar1}, in which the process 
$e^+ e^- \to K^+ K^- \pi^0 $ was studied using the initial state radiation
method. In Ref.~\cite{bib:babar1}, it is shown that the process 
$e^+ e^- \to K^+ K^- \pi^0$ proceeds through the $K^{\ast\pm}(892) K^{\mp}$,
$ \phi (1020) \pi^0 $, and $K^{\ast \pm}_2(1430) K^\mp $ intermediate
states.  In the VEPP-2000 energy range, $\sqrt {s} <2 $ GeV,
the $K^{\ast\pm}_2(1430) K^\mp $ contribution is expected to be small.
The cross section of the process $e^+ e^- \to \phi(1020) \pi^0 $
was also measured in the BABAR experiment~\cite{bib:babar2} in the final
state $K_SK_L\pi^0$.

The aim of this work is to measure the cross section for the process
$e^+ e^- \to K^+ K^- \pi^0 $ with an accuracy comparable to that 
of BABAR~\cite{bib:babar1}.

\section {Detector and experiment\label{sec:det}}
The VEPP-2000 $e^+e^-$ collider operate in the c.m. energy range from 0.32 to
2.01 GeV. SND~\cite{bib:snd} is a general-purpose non-magnetic detector. It
comprises a tracking system, a particle identification system
based on aerogel threshold Cherenkov counters, an electromagnetic calorimeter,
and a muon system. The main part of the detector is a three-layer spherical
calorimeter based on NaI (Tl) crystals with a thickness of 13.4$X_0$, where
$X_0$ is the radiation length. Its energy resolution is
$\sigma_{E_{\gamma}}/E_{\gamma} = 4.2\%/\sqrt[4]{E_{\gamma}~\textrm {(GeV)}}$,
and the angular resolution is
$\sigma_{\theta,\phi}=0.82^{\circ}/\sqrt{E_{\gamma}(\mbox{GeV})}$,
where $E_\gamma$ is the photon energy.
The calorimeter covers about 95\% of the solid angle.

The tracking system, which is used for measurement of directions and
production points of charged particles, is located inside the calorimeter, 
around the collider beam pipe. It consists of a nine-layer cylindrical drift
chamber and a proportional chamber with cathode strip readout. The tracking
system covers a solid angle of 94\% of $4\pi$.

The charged particle identification is provided by the system of
aerogel Cherenkov counters (ACC)~\cite{bib:kk}. It consists of nine counters
forming a cylinder located around the tracking system. The counters cover
the polar angle region $50^\circ<\theta<132^\circ$. The aerogel
radiator has a refractive index of $n=1.13$ and a thickness of 30~mm.
The Cherenkov light is collected and transmitted to photodetectors using
wavelength shifters located inside the aerogel radiator. Information from 
the ACC is used only if the charged particle track extrapolates
to the ACC active area that excludes the regions of shifters and gaps 
between counters. The active area is 81\% of the ACC area.

The calorimeter is surrounded by the 10 cm thick iron absorber and 
the muon system, which consists of a layer
proportional tubes and a layer of scintillation counters with
an 1~cm thick iron sheet between them.

In this work we analyze a data sample with an integrated luminosity of
26.4~pb$^{-1}$ recorded in 2011--2012. In the energy range under study,
1.27--2.00 GeV, data were collected in 44 energy points. Because of the absence
of narrow structures in the cross sections under study, these energy points
are merged into 27 energy intervals. The luminosity-weighted average 
c.m. energies for these intervals are listed in Table~\ref{table1}.

For simulation of signal events, a Monte Carlo (MC) event generator
is used based on formulas from Ref.~\cite{wppp}.
It is assumed that the process $e^+e^-\to K^+K^-\pi^0$ proceeds through the 
$K^{\ast}(892)^{\pm} K^{\mp}$ and $\phi\pi^0 $ intermediate states.
The following background processes are also simulated:
\begin{eqnarray}
e^+e^-&\to&\pi^+\pi^-\pi^0,\, \pi^+\pi^-\pi^0\pi^0,\,
\pi^+\pi^-\pi^0\pi^0\pi^0,\nonumber \\
e^+e^-&\to&K^+ K^-,\,K_S K_L,\,K_S K^{\pm}\pi^{\mp},\,K_LK^{\pm}\pi^{\mp},\nonumber \\
e^+e^-&\to& K^+ K^- \pi^0 \pi^0,\,K_S K^{\pm}\pi^{\mp}\pi^0,\,
K_L K^{\pm}\pi^{\mp}\pi^0.\label{bkg_proc}
\end{eqnarray}
Event generators for the signal and background processes include radiation
corrections~\cite{RadCor}. The angular distribution of the extra photon
emitted from the initial state is generated according to Ref.~\cite{BM}.
Interactions of the generated particles with the detector materials
are simulated using the GEANT4 software~\cite{bib:geant}.
The simulation takes into account variations of experimental 
conditions during data taking, in particular, dead detector channels, 
and beam-generated background. The beam background leads to the appearance
of spurious charged tracks and photons in the events of interest. To take
this effect into account, the simulation uses special background events
recorded during data taking with a random trigger, which are superimposed
on simulated events.

The integrated luminosity is measured on $e^+ e^- \to e^+ e^-$ events 
with an uncertainty better than 2\%~\cite{bib:epp2k}.

\section { Events selection\label{sec:sel}}
Events from the $e^+ e^- \to K^+ K^- \pi^0$ process are detected as
two charged particles and two photons from the $\pi^0$ decay.
An event may contain additional charged tracks originating from $\delta$
electrons and beam background, and spurious photons originating from 
splitting of the electromagnetic shower, kaon nuclear interaction in the 
calorimeter, and beam background.
We select events with two or three charged particles and two or more photons
with energy higher than 30 MeV. The charged-particle track is required to have
at least 4 hits in the drift chamber. At least two charged particles must 
originate from the interaction region, i.e. satisfy the conditions:
$d_i < 0.3$~cm, $| z_i| <10$~cm, $i = 1,2$, and $|z_1-z_2|<5$~cm, where $d_i$
is the distance between the track and the beams axis, and the $z_i$ is 
the $z$-coordinate of the track point closest to the beam axis. If there are 
three charged particles satisfying the above criteria, two of them with the 
best $\chi^2$ of the fit to a common vertex are selected. The third must have
$d_3 > 0.2$~cm.

\begin{figure}
\centering
\resizebox{0.45\textwidth}{!}{\includegraphics{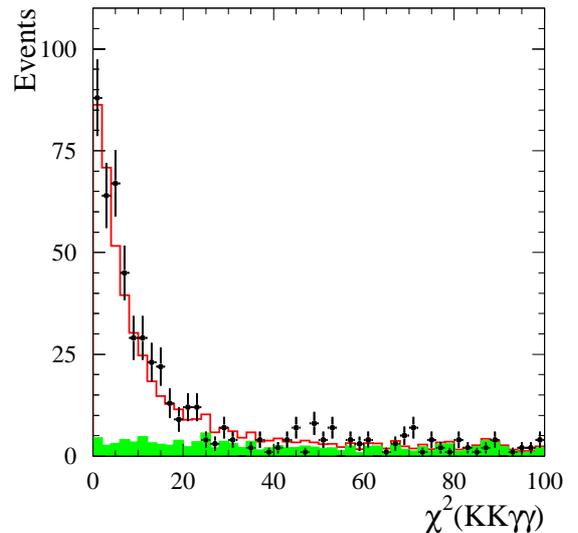}}
\caption{\label{chi2kk}
The $\chi^2(KK2\gamma)$ distribution for data events with $100 \le
m_{\gamma\gamma} \le 170$~MeV/$c^2$ from the interval
$1.5<\sqrt{s}<1.72$ GeV (points with error bars).
The solid histogram is the sum of the simulated signal distribution and the
background distribution. The hatched histogram represents the background.}
\end{figure}
For events passing the primary selection described above, the kinematic fit 
with four constraints of energy and momentum balance to the hypothesis
$e^+ e^- \to K^+ K^- \gamma \gamma$ is performed. From the fit, we determine
the kaon momenta and refine the photon energies. The quality of the fit is
characterized by the parameter $\chi^2 (KK2\gamma)$. If there are more than
two photons in an event, all two-photon combinations are tested and one with
the smallest $\chi^2$ is selected. The $\chi^2 (KK2\gamma)$ distributions for
signal and background events are shown in Fig.~\ref{chi2kk}. A method to
obtain the background distribution is described in Sec.~\ref{sec:eff}. The 
fitted photon parameters are used to calculate the two-photon invariant mass
$m_{\gamma\gamma}$. The kinematic fits are also performed 
to the hypotheses  $\pi^+ \pi^- \gamma \gamma$ and
$\pi^+ \pi^- \pi^0 \pi^0 $, and the parameters $\chi^2(2\pi 2\gamma)$ and
$\chi^2(4\pi)$ are determined. The fit to the $\pi^+ \pi^- \pi^0 \pi^0$ 
hypothesis includes two additional $\pi^0$-mass constraints and is applied 
to events with four photons.
To select the events of the process
$e^+ e^- \to K^+ K^- \pi^0$, the following conditions are used:
\begin{eqnarray}
\chi^2 (KK2\gamma)& < &40,\nonumber \\
\chi^2 (2\pi2\gamma)& > &20,\nonumber \\
\chi^2 (4\pi)& > &20.\nonumber
\end{eqnarray}

\section {Kaon identification\label{pid}}
\begin{figure}[!htbp]
\centering
\resizebox{0.48\textwidth}{!}{\includegraphics{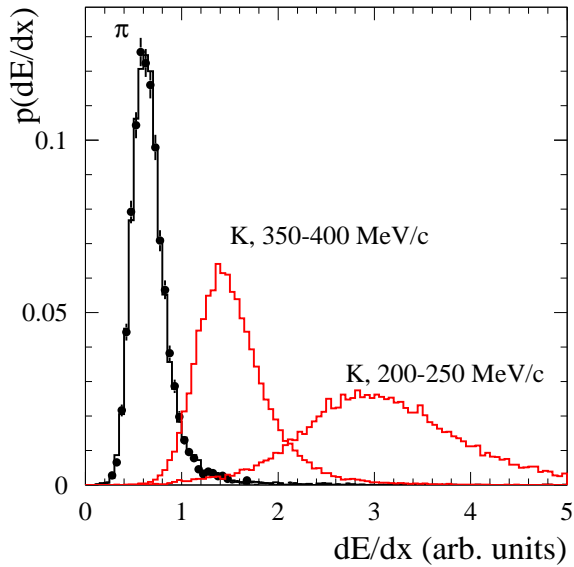}}
\caption{\label{dedxpi}
The probability density distribution of the ionization losses
in the drift chamber for pions and kaons. The points with
error bars represent the data distribution for pions from 
$e^+ e^- \to \pi^+\pi^- \pi^0 \pi^0 $ events, the histogram is the same
simulated distribution. The kaon distributions for two momentum ranges are
obtained using $e^+e^-\to K^+K^-\pi^0$ simulation.}
\end{figure}
For kaon identification, information about ACC response and
ionization losses of charged particles in the drift chamber ($dE/dx$)
measured in e$^{\pm}$ dE/dx units is used.

In the energy range of VEPP-2000 charged kaons do not produce a Cherenkov
signal in the ACC. For pions the threshold momentum is 265 MeV/$c$.

The $dE/dx$ distribution for pions from the background process
$e^+ e^- \to \pi^+ \pi^- \pi^0 \pi^0 $ is shown in Fig.~\ref{dedxpi}.
For kaons from the process $e^+e^-\to K^+K^-\pi^0$ in the energy range under
study, momenta vary from 100 MeV/$c$ to 800 MeV/$c$, and there is a
strong dependence of $dE/dx$ on the kaon momentum. It is illustrated
in Fig.~\ref{dedxpi}, where the kaon $dE/dx$ distributions obtained
using $e^+e^-\to K^+K^-\pi^0$ simulation is shown for two ranges of kaon
momentum.
\begin{figure*}[!htbp]
\centering
\resizebox{0.98\textwidth}{!}{\includegraphics{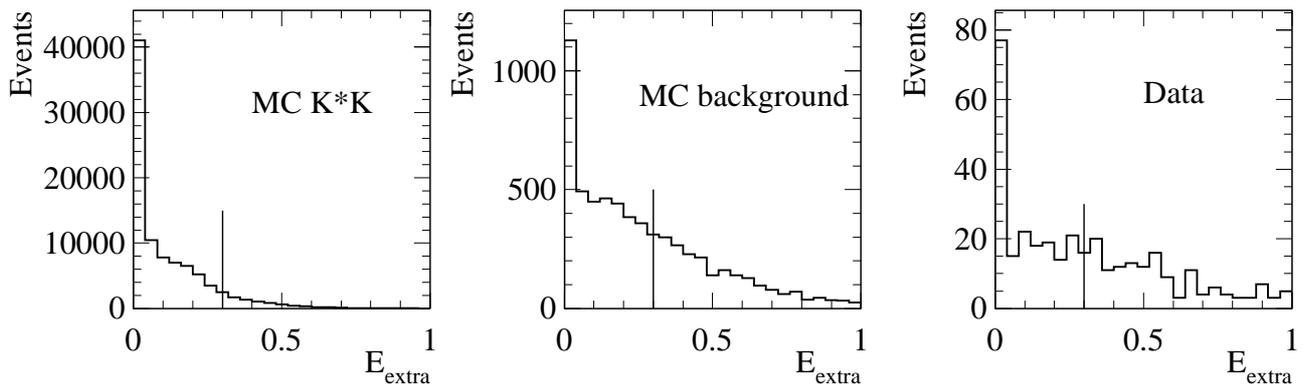}}
\caption{\label{sumegn}
The $E_{\rm extra}$ distribution for data events with
$\sqrt{s}>1.8$~GeV and simulated events of the process under study and
background processes. The vertical line indicates
the boundary of the condition $E_{\rm extra}<0.3$.}
\end{figure*}
\begin{figure*}[!htbp]
\centering
\resizebox{0.98\textwidth}{!}{\includegraphics{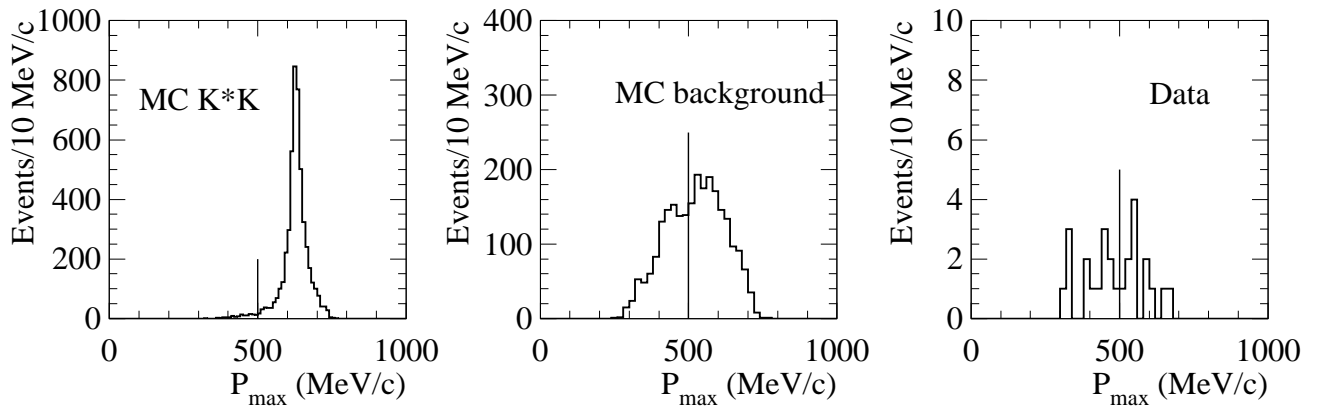}}
\caption{\label{maxmom}
The $P_{\rm max}$ distribution for selected data events and simulated
events of the process under study and background processes at 
$\sqrt{s}=1.89$~GeV. The vertical line indicates
the boundary of the condition $P_{\rm max}>500$~MeV/$c$.}
\end{figure*}

A charged particle is identified as a kaon if it passes through the
active ACC area and does not produce a Che\-ren\-kov signal. If the
momentum of this particle determined from the kinematic fit to the
$e^+ e^- \to K^+ K^- \pi^0$ model is less than 300 MeV/$c$,
the additional condition $dE/dx>1$ is applied.
We select events with one or two identified kaons. For
events with one identified kaon, the second charged particle must not
pass the active ACC region, have the polar angle in the range from 
$ 40^\circ$ to $ 140^\circ$, the fitted momentum less than
450~MeV/$c$, and $dE/dx >1$.

\section {Background suppression\label{sec:bkg}}
The significant background for the process under study comes from  multihadron
processes containing several neutral pions in the final state.
To suppress this background, the condition $E_{\rm extra}<0.3$ is used,
where $E_{\rm extra}$ is the total energy of photons not included in the
kinematic fit, normalized to the beam energy $\sqrt{s}/2$. 
The $E_{\rm extra}$ distributions for selected data events, signal simulation,
and simulation of the background processes~(\ref{bkg_proc})
are shown in Fig.~\ref{sumegn}, for $\sqrt{s}>1.8$~GeV, where the effect of
the cut on $E_{\rm extra}$ is maximal. The contributions of different
background processes to the background spectrum are calculated using their 
measured cross sections. 

For additional suppression of background, the conditions
on the minimum ($P_{\rm min}$) and maximum ($P_{\rm max}$) kaon momenta
in an event obtained from the kinematic fit to the 
$e^+ e^- \to K^+ K^- \gamma \gamma$ hypothesis are used.
The minimum kaon momentum is required to be larger than 100 MeV/$c$, while
the cut on the maximum momentum depends on c.m. energy and is chosen such
that the fraction rejected signal events does not exceed 10\%.
Figure~\ref{maxmom} shows the $P_{\rm max}$ distribution 
for selected data events at $\sqrt{s}=1.89$ GeV, and the simulated
distributions for the process under study and background processes.
At this energy, $P_{\rm max}>500$~MeV/$c$ is required.

To suppress the background from collinear events of the processes
$e^+ e^- \to e^+ e^- $, $\pi^+ \pi^-$, $ K^+ K^-$, we reject events 
with $|\Delta \varphi|<5^{\circ}$ and $|\Delta \theta|<5^{\circ}$, 
where $\Delta \varphi=|\varphi_1-\varphi_2|-180^\circ$,
$\Delta \theta=\theta_1+\theta_2-180^\circ$, and $\varphi_i$ and
$\theta_i$ are the azimuthal and polar angles of the charged particles,
respectively.

The process $ e^+ e^- \to \phi \pi^0 $ will be analyzed
separately in Sec.~\ref{sec:phipi0}. When studying the $e^+e^-\to K^+K^-\pi^0$
process, the $\phi \pi^0$ events are removed by the condition
$m_{\rm rec}>1.05$~GeV/$c^2$,
where $m_{rec}$ is the mass recoiling against the photon pair
calculated after the kinematic fit
to the $e^+ e^- \to K^+ K^- \gamma \gamma$ hypothesis.

\begin{figure}[!htbp]
\centering
\resizebox{0.45\textwidth}{!}{\includegraphics{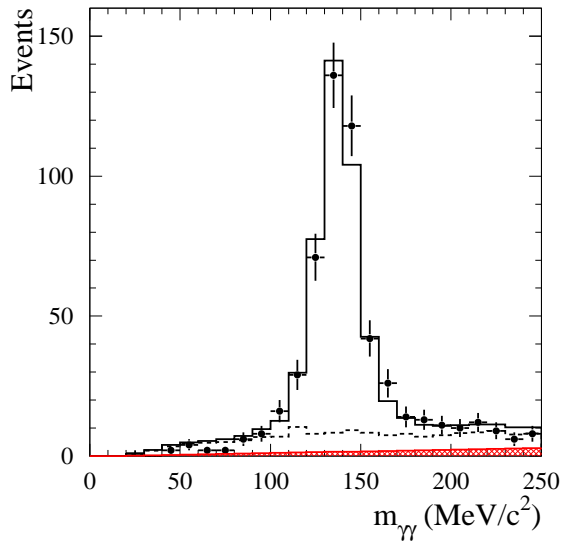}}
\caption{\label{m12k800}
The two-photon invariant mass spectrum for selected data events with
$\sqrt{s}=1.5$--1.72~GeV (points with errors).
The solid histogram is the result of the fit to the data spectrum with
the sum of the signal and background distributions. The dashed histogram
represents the fitted background. The hatched histogram shows 
the part of the background described by the linear function.}
\end{figure}

The two-photon invariant mass spectrum for selected data events from the
energy range $\sqrt{s}=1.50$--1.72~GeV, where the $e^+e^-\to K^+K^-\pi^0$ 
cross section is maximal, is shown in Fig.~\ref{m12k800}. This
spectrum in the mass range $30 <m_{\gamma \gamma} <250$ MeV/$c^2$ is fitted by 
a sum of signal and background distributions. The signal distribution 
is obtained using the $e^+e^-\to K^+K^-\pi^0$ simulation. The background 
distribution is a sum of the simulated mass spectrum for the 
processes~(\ref{bkg_proc}) and a linear function
describing contribution of other background processes. The simulated
background spectrum is multiplied by the scale factor $\alpha_{\rm b}$.
During the fit, $\alpha_{\rm b}$ is varied 
within 10\% around unity. The fit result is shown in Fig.~\ref{m12k800} by
the solid histogram. The dashed histogram represents the total fitted 
background. The hatched histogram shows the part of the background described 
by the linear function. It is seen that the background 
processes~(\ref{bkg_proc}) describe approximately 80\% of the background 
observed in data.

To estimate the systematic uncertainty in the number of signal events 
due to incorrect description of the background shape, the fit with
free $\alpha_{\rm b}$ is performed. The difference between the results of the
two fits  is taken as a measure of systematic uncertainty. The fitted
numbers of $e^+e^-\to K^+K^-\pi^0$ events with the statistical and systematic
uncertainties for different energy points are listed in Table~\ref{table1}. 
In the energy range $\sqrt{s}=1.45$--1.70~GeV the 
systematic uncertainty is about 5\%.
\begin{figure}[!htbp]
\centering
\resizebox{0.45\textwidth}{!}{\includegraphics{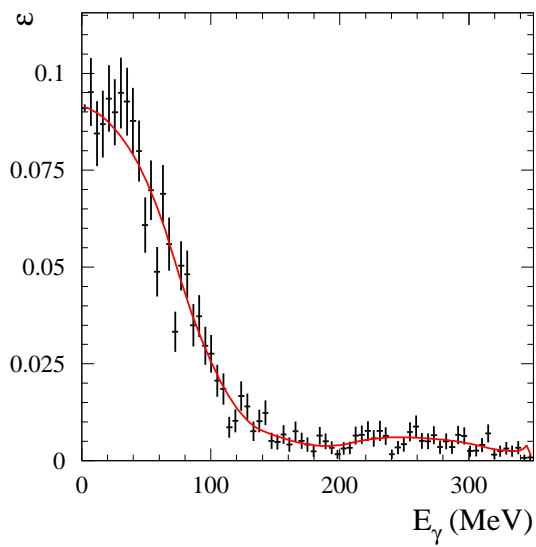}}
\caption{\label{effg787}
The dependence of the detection efficiency for $e^+e^-\to K^+K^-\pi^0$ events at 
$\sqrt{s}=1.575$~GeV  on the energy of the photon emitted from the initial
state. The dependence is approximated by a smooth function.}
\end{figure}

\section{Detection efficiency\label{sec:eff}}
The visible cross for the process under study 
$\sigma_{{\rm vis},i}={N_i}/{L_i}$, where $N_i$ and $L_i$ are the number of 
selected events and the integrated luminosity for the $i-$th energy point, 
is related to the Born cross section $\sigma_{0}$
by the following expression:
\begin {equation}
\sigma_{\rm vis}(\sqrt{s})=\int\limits^{z_{\rm max}}_{0} dz\sigma_{0}
(\sqrt{s(1-z)})F(z,s)\varepsilon(\sqrt{s},z),
\label {bornsec}
\end {equation}
where $F(z,s)$ is a function describing the probability of emission of
photons with the energy
$z\sqrt {s}/2$ from the initial state~\cite{RadCor},
$\varepsilon(\sqrt{s},z)$ is the detection efficiency,
$z_{\rm max}=1-(m_{\pi^0}+2m_K)^2/s$, $m_{\pi^0}$ and $m_K$ are the
$\pi^0$ and $K^\pm$ masses, respectively.

The detection efficiency for $e^+e^-\to K^+K^-\pi^0$ events is determined
using MC simulation as a function of $\sqrt{s}$ and $z$. The dependence of the
efficiency on $z$ at $\sqrt{s}=1.575$ GeV is shown in Fig.~\ref{effg787}.
The values of the efficiency at zero photon energy
$\varepsilon_0(\sqrt{s})= \varepsilon (\sqrt{s},0)$ for different energy
points are listed in Table~\ref{table1}.

Inaccuracy in simulation of distributions of parameters used in
event selection leads to a systematic uncertainty in the detection
efficiency determined using the simulation. The most critical 
selection parameters are $\chi^2(KK\gamma\gamma)$, $dE/dx$, and
$E_{\rm extra}$. To estimate the systematic uncertainty, we use
events from the energy region $1.5 <\sqrt{s}<1.72$~GeV, where the 
$e^+e^-\to K^+K^-\pi^0$ cross section is maximal, change the selection conditions,
and study the change in the measured signal cross section.
For the parameters mentioned above, the loosened selection criteria
$\chi^2(KK\gamma\gamma)<80 $, $dE/dx>0.8$, and $E_{\rm extra}<0.5$ are
used instead of the standard criteria $\chi^2(KK\gamma\gamma)<40$, 
$dE/dx>1$, and $E_{\rm extra}<0.3$.  It is found that the total systematic
uncertainty due to these conditions does not exceed 8$\%$.

Figure~\ref{chi2kk} shows the $\chi^2(KK2\gamma)$ distribution for data events
with $100 \le m_{\gamma\gamma} \le 170$~MeV/$c^2$ from the interval 
$1.5<\sqrt{s}<1.72$~GeV.
It is seen that the data distribution is in good agreement with the sum of 
the simulated signal distribution and the background distribution.
The latter is a sum of the simulated distribution for the background
processes~(\ref{bkg_proc}) and the distribution for unaccounted background,
which fraction is about 20\% (see Sec.~\ref{sec:bkg}). We assume that this 
unaccounted background has a linear shape of the $m_{\gamma\gamma}$ spectrum 
and therefore can be estimated in each $\chi^2$ bin using the equation 
$N_{lin}=(N_2-r_sN_1)/(2-r_s)$, where $N_1$ and $N_2$ are the numbers of 
selected data events with subtracted background from the processes (1) in the
signal region ($100 < m_{\gamma\gamma} < 170$ MeV/$c^2$) and the sidebands
($30 < m_{\gamma\gamma} < 100$ MeV/$c^2$ and $170 < m_{\gamma\gamma} < 240$
MeV/$c^2$), respectively, and $r_s$ is the $N_2/N_1$ ratio for signal events
obtained using simulation.

Other sources of the systematic uncertainty on the detection efficiency 
were studied in Ref.~\cite{bib:sndkk}. These are the uncertainties associated
with the kaon identification using the ACC (1.2\%), the definition 
of the ACC active region (0.3\%), the inaccuracy in simulation of kaons 
nuclear interaction (0.1\%), and the photon conversion in material before
the tracking system (0.7\%). The total systematic uncertainty on 
the detection efficiency is 8\%.

\section{\boldmath Study of the $K^\pm\pi^0$ invariant mass spectrum}
\begin{figure}
\centering
\resizebox{0.45\textwidth}{!}{\includegraphics{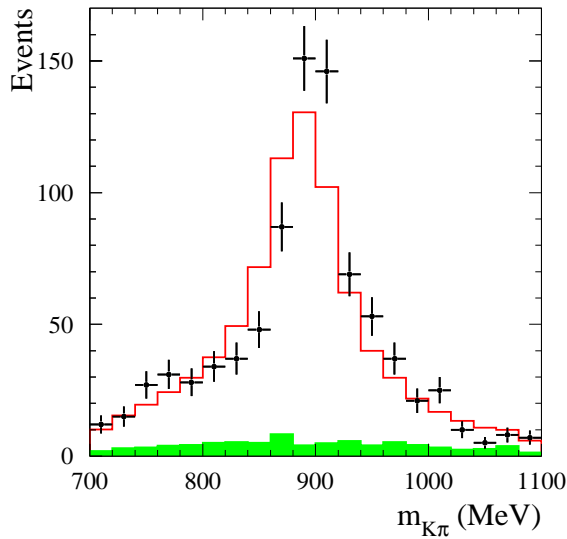}}
\caption{\label{invmkpi}
The $K\pi^0$ invariant mass spectrum for data events from the energy
range $1.5<\sqrt {s}<1.72$~GeV (points with error bars). The solid histogram
is the sum of the simulated $e^+e^-\to K^+K^-\pi^0$ distribution and
background. The hatched histogram represents the background.}
\end{figure}
It is shown in Ref.~\cite{bib:babar1} that the process
$e^+ e^- \to K^+ K^- \pi^0$ proceeds through the $K^{\ast\pm}(892)K^{\mp}$ and
$K^{\ast\pm}_2(1430) K^{\mp}$ intermediate states. In the VEPP-2000 energy 
range, below 2 GeV, the dominant intermediate state is expected to be 
$K^{\ast\pm}(892) K^{\mp}$.
Figure~\ref{invmkpi} shows the $K\pi^0$ invariant mass spectrum for
data events from the energy region $1.5<\sqrt {s}<1.72$~GeV.
The background contribution is estimated in the same way as for the
$\chi^2(KK2\gamma)$ distribution in Sec.~\ref{sec:eff}. The solid histogram in
Fig.~\ref{invmkpi} represents the signal plus background distribution. The 
signal $K\pi^0$ mass spectrum is obtained using the simulation in the model
$e^+ e^- \to K^{\ast\pm}(892)K^{\mp}\to K^+ K^- \pi^0$. It is seen that the 
$K^{\ast\pm}(892)K^{\mp}$ intermediate state is dominant in the 
$e^+e^- \to K^+ K^- \pi^0$ reaction. The observed difference between data and
simulated distributions may be due to a contribution from other intermediate
states, e.g. $\phi\pi^0$, $K^{\ast\pm}(1410)K^{\mp}$, and
$K^{\ast\pm}_2(1430) K^{\mp}$.  Their interference with the dominant 
$K^\ast(892)K$ amplitude may lead to a shift and narrowing of the
$K^\ast(892)$ peak in Fig.~\ref{invmkpi}.

From Fig.~\ref{invmkpi} we roughly estimate that the contribution of 
intermediate states other than $K^\ast(892)\pi$ does not exceed 20\%.
The difference in the detection efficiency between different intermediate 
state is estimated comparing the efficiencies for simulated $K^\ast(892)K$
events and $\phi\pi^0$ events with $K^+K^-$ invariant mass higher than 
1.04 GeV/$c^2$. This difference does less than 20\%. So, we estimate
that the model uncertainty in the detection efficiency due to the
contribution of non-$K^\ast(892)K$ intermediate states does not exceed 4\%.

\section{\boldmath Born cross section for the process 
$e^+e^- \to K^+ K^- \pi^0$\label{kstkfit}}
The formula (\ref{bornsec}) given in Sec.~\ref{sec:eff} describes
the relation between the visible and Born cross sections.  The experimental
values of the Born cross section are determined 
in the following way. The measured energy dependence of the visible cross
section is approximated by Eq.~(\ref{bornsec}), in which the Born
cross section is parametrized by some model that describes
data reasonably well. As a result of the approximation, model parameters are
determined and the radiation corrections are calculated as
$1+\delta(s)=\sigma_{\rm vis}(s)/(\varepsilon_0(s)\sigma_{0}(s))$.
The experimental value of the Born cross section is then determined as
\begin{equation}
\sigma_{0,i}=\frac{\sigma_{{\rm
vis},i}}{\varepsilon_0(\sqrt{s_i})(1+\delta(s_i))}.
\label{secborni}
\end{equation}

In Ref.~\cite{bib:babar1}, the isoscalar and isovector cross sections for the
process $e^+e^- \to K^{\ast}K$ were measured separately, and it was shown
that the isoscalar amplitude dominates only near the maximum of the 
$\phi(1680)$ resonance. Below 1.55 GeV and above 1.8 GeV the isoscalar and
isovector amplitudes are of the same order of magnitude.
In the current analysis, a simplified two-resonance model is used to describe
the $e^+e^-\to K^+K^-\pi^0$ Born cross section:
\begin{eqnarray}
\sigma_0(\sqrt{s})&=&\left|
\frac{A_0 M_0\Gamma_0}{M^2_0-s-i\sqrt{s}\Gamma_0}\right.\nonumber \\
&+&\left.
\frac{A_1e^{i\psi}M_1\Gamma_1}{M^2_1-s-i\sqrt{s}\Gamma_1} \right|^2
\frac{P(s)}{s^{3/2}},
\label{eq9}
\end{eqnarray}
where $M_i$ and $\Gamma_i$ are the masses and widths of two effective
resonances, $A_i$ are their real amplitudes, and $\psi$ is
the relative phase between the amplitudes. The function $P(s)$ describes
the energy dependence of the $K^{\ast\pm}(892)K^\mp$ phase space, which
takes into account the finite $K^\ast(892)$ width and
the interference of the $K^{\ast +}K^-$ and  $K^{\ast -}K^+$
amplitudes.
In this model, the first term in Eq.~(\ref{eq9}) describes the total 
contribution of the low-lying resonances $\rho(770)$, $\omega(782)$, and 
$\phi(1020)$, and the excitations $\rho(1450)$ and $\omega(1420)$. The 
parameters $M_0$ and $\Gamma_0$ are taken to be equal to the mass and width of
the $\phi(1020)$.
The second term describes the total contribution of all excited vector
resonances. Parameters $A_0$, $A_1$, $M_1$, $\Gamma_1$ and $\psi$
are determined from the fit to the visible cross section data.

\begin{figure}
\centering
\resizebox{0.45\textwidth}{!}{\includegraphics{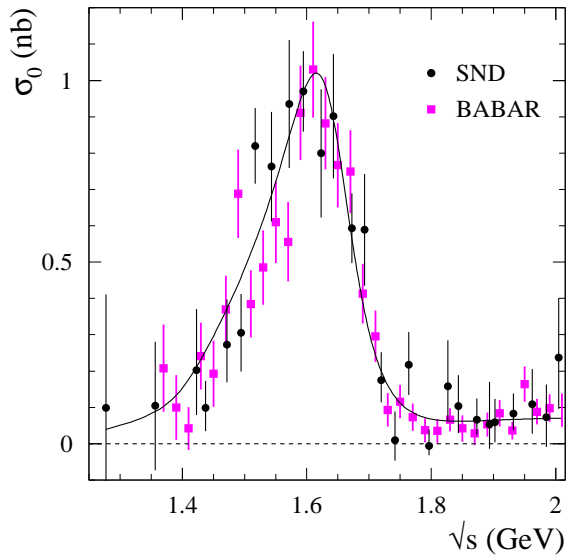}}
\caption{\label{fig_crs}
The $e^+e^-\to K^+K^-\pi^0$ Born cross section measured in this work
(circles) compared with the BABAR~\cite{bib:babar1} data (squares).
The curve is the result of fit described in the text.
}
\end{figure}

The values of the Born cross section calculated using Eq.(\ref{secborni})
and the fitted curve are shown in Fig.~\ref{fig_crs}. The model describes
the data reasonably well: $\chi^2/{\rm ndf}=28.2/22$, where ${\rm ndf}$
is the number of degrees of freedom ($P(\chi^2)=16.9\%$).
The fitted values of the mass and width, $M_1=1662 \pm 20$~MeV/$c^2$,
$\Gamma_1=159 \pm 32$~MeV, are close to the Particle Data Group (PDG) 
values for the $\phi(1680)$ resonance~\cite{pdg}, indicating that 
this resonance dominates the $e^+e^-\to K^+K^-\pi^0$ cross section.

The obtained values of the radiation correction and
Born cross section are listed in Table~\ref{table1}. For the cross
section, the statistical and energy dependent systematic uncertainties
are quoted. The latter includes the systematic uncertainty in the
number of $e^+e^-\to K^+K^-\pi^0$ events, and the model error of radiation
correction, which is determined by varying the model parameters obtained in
fit within their errors. The energy independent correlated systematic 
uncertainty is 9\%. It includes the systematic uncertainties in the luminosity
measurement (2\%) and detection efficiency (8\%), and the model error
of the detection efficiency (4\%).
\begin{table*}
\centering
\caption{\label{table1}
The c.m. energy ($\sqrt{s}$), integrated luminosity ($L$),
%number of selected  $e^+ e^- \to K^+ K^- \pi^0$ events ($N$),
fitted number of events of $e^+e^- \to K^+K^-\pi^0$ process (N)
detection efficiency ($\varepsilon_0$), 
radiation correction factor ($1+\delta$),
and Born cross section for the process $e^+ e^- \to
K^+ K^- \pi^0$ ($\sigma_0$). For the number of events,
statistical and systematic errors are quoted. For the cross section, 
the second error is the energy-dependent uncorrelated systematic uncertainty.
The energy independent correlated uncertainty on the cross section is 9\%.
}
\begin{tabular}{cccccc}
\hline
$\sqrt{s}$ (GeV) & $L$ (nb$^{-1}$) &$N$ &
$\varepsilon_0$ &$1+\delta$ &$\sigma_0$ (nb) \\
\hline
$ 1.277 $ & $  763 $ & $  0.7^{+ 2.1 }_{- 2.1 } \pm 1.3$ & $ 0.011 $ &
$0.810$ & $ 0.099^{+ 0.315}_{- 0.311} \pm 0.189$\\
$ 1.357 $ & $  845 $ & $  1.5^{+ 2.6 }_{- 2.6 } \pm 0.6$ & $ 0.020 $ & $0.874$ & $ 0.105^{+ 0.177}_{- 0.178} \pm 0.043$\\
$ 1.423 $ & $  588 $ & $  3.4^{+ 2.8 }_{- 2.1 } \pm 1.0$ & $ 0.035 $ & $0.817$ & $ 0.203^{+ 0.167}_{- 0.124} \pm 0.061$\\
$ 1.438 $ & $ 1505 $ & $  5.5^{+ 4.2 }_{- 3.6 } \pm 1.7$ & $ 0.045 $ & $0.823$ & $ 0.098^{+ 0.075}_{- 0.064} \pm 0.032$\\
$ 1.471 $ & $  619 $ & $  9.1^{+ 4.1 }_{- 3.4 } \pm 0.7$ & $ 0.064 $ & $0.840$ & $ 0.273^{+ 0.125}_{- 0.103} \pm 0.022$\\
$ 1.494 $ & $  754 $ & $ 14.4\pm 5.0           \pm 0.1$ & $ 0.075 $ & $0.831$ & $ 0.306\pm 0.107	    \pm 0.002$\\
$ 1.517 $ & $ 1448 $ & $ 83.2\pm10.6           \pm 4.3$ & $ 0.083 $ & $0.845$ & $ 0.820\pm 0.104	    \pm 0.042$\\
$ 1.543 $ & $  578 $ & $ 33.0\pm 6.5           \pm 0.1$ & $ 0.088 $ & $0.846$ & $ 0.763\pm 0.151	    \pm 0.003$\\
$ 1.572 $ & $  533 $ & $ 39.0\pm 7.3           \pm 0.7$ & $ 0.091 $ & $0.857$ & $ 0.936\pm 0.176	    \pm 0.015$\\
$ 1.595 $ & $ 1284 $ & $ 94.7\pm10.7           \pm 5.6$ & $ 0.087 $ & $0.873$ & $ 0.970\pm 0.110	    \pm 0.050$\\
$ 1.623 $ & $  545 $ & $ 34.2\pm 7.5           \pm 2.6$ & $ 0.089 $ & $0.885$ & $ 0.800\pm 0.176	    \pm 0.060$\\
$ 1.643 $ & $  499 $ & $ 33.0\pm 6.2           \pm 4.2$ & $ 0.081 $ & $0.907$ & $ 0.902\pm 0.171	    \pm 0.118$\\
$ 1.672 $ & $ 1397 $ & $ 59.1\pm 9.6           \pm 2.7$ & $ 0.071 $ & $1.011$ & $ 0.593\pm 0.096	    \pm 0.032$\\
$ 1.693 $ & $  490 $ & $ 19.2\pm 5.0           \pm 2.3$ & $ 0.063 $ & $1.053$ & $ 0.589\pm 0.154	    \pm 0.077$\\
$ 1.720 $ & $ 1051 $ & $ 13.1^{+ 5.8}_{- 4.6 } \pm 1.5$ & $ 0.060 $ & $1.190$ & $ 0.174^{+ 0.078}_{- 0.061} \pm 0.022$\\
$ 1.742 $ & $  529 $ & $  0.4^{+ 2.9 }_{- 2.1 } \pm 0.6$ & $ 0.057 $ & $1.229$ & $ 0.010^{+ 0.079}_{- 0.057} \pm 0.016$\\
$ 1.764 $ & $ 1290 $ & $ 16.0^{+ 6.6}_{- 6.1 } \pm 0.8$ & $ 0.048 $ & $1.178$ & $ 0.218^{+ 0.090}_{- 0.083} \pm 0.003$\\
$ 1.797 $ & $ 1424 $ & $ -0.5^{+ 3.5 }_{- 2.0 } \pm 0.6$ & $ 0.052 $ & $1.054$ & $-0.006^{+ 0.045}_{- 0.026} \pm 0.007$\\
$ 1.826 $ & $  529 $ & $  4.2^{+ 3.3 }_{- 2.6 } \pm 2.2$ & $ 0.047 $ & $1.071$ & $ 0.158^{+ 0.126}_{- 0.096} \pm 0.097$\\
$ 1.844 $ & $ 1006 $ & $  5.1^{+ 4.2 }_{- 3.7 } \pm 3.7$ & $ 0.048 $ & $1.028$ & $ 0.104^{+ 0.085}_{- 0.074} \pm 0.084$\\
$ 1.873 $ & $ 1606 $ & $  4.8^{+ 4.2 }_{- 3.1 } \pm 0.0$ & $ 0.047 $ & $0.964$ & $ 0.066^{+ 0.058}_{- 0.043} \pm 0.027$\\
$ 1.893 $ & $  624 $ & $  1.4^{+ 3.1 }_{- 1.8 } \pm 0.6$ & $ 0.046 $ & $0.937$ & $ 0.053^{+ 0.117}_{- 0.068} \pm 0.022$\\
$ 1.903 $ & $ 1456 $ & $  3.7^{+ 4.0 }_{- 3.5 } \pm 4.0$ & $ 0.045 $ & $0.956$ & $ 0.059^{+ 0.065}_{- 0.056} \pm 0.068$\\
$ 1.932 $ & $ 2235 $ & $  7.0^{+ 4.7 }_{- 3.8 } \pm 5.8$ & $ 0.042 $ & $0.907$ & $ 0.083^{+ 0.056}_{- 0.046} \pm 0.071$\\
$ 1.962 $ & $  971 $ & $  3.8^{+ 3.4 }_{- 2.8 } \pm 1.5$ & $ 0.039 $ & $0.913$ & $ 0.109^{+ 0.097}_{- 0.081} \pm 0.044$\\
$ 1.985 $ & $ 1204 $ & $  3.2^{+ 4.0 }_{- 3.6 } \pm 0.3$ & $ 0.039 $ & $0.942$ & $ 0.073^{+ 0.090}_{- 0.083} \pm 0.069$\\
$ 2.006 $ & $  582 $ & $  4.9^{+ 3.4 }_{- 2.9 } \pm 1.0$ & $ 0.037 $ & $0.956$ & $ 0.238^{+ 0.164}_{- 0.142} \pm 0.047$\\
\hline
\end{tabular}
\end{table*}

In Fig.~\ref{fig_crs}, our measurement of the  $e^+ e^- \to K^+ K^-
\pi^0$ cross section  is compared with the result of the most precise
previous measurement by BABAR~\cite{bib:babar1}. Two measurements are
consistent and comparable in accuracy.

\section{\boldmath Study of the process 
$e^+ e^- \to \phi \pi^0 \to K^+ K^- \pi^0$\label{sec:phipi0}}
The selection criteria for $e^+ e^- \to \phi \pi^0 \to K^+ K^- \pi^0$
events are close to those described in Sec~\ref{sec:sel}. Events with  mass
recoiling against the photon pair $m_{\rm rec}<1.08$ GeV/$c^2$ are analyzed.
The requirements on the minimum and maximum momenta of charged kaons are
removed. To suppress background from the initial state radiation process
$e^+ e^- \to \phi(1020) \gamma \to K^+ K^- \gamma$,
the additional condition is imposed that the difference between the normalized
energy of the most energetic photon in event $2E_{\gamma,max}/\sqrt{s}$ and
$(1-M_\phi^2/s)$ is larger than 0.1. Here $M_\phi$ is the $\phi(1020)$
mass.

\begin{figure*}
\centering
\resizebox{0.7\textwidth}{!}{\includegraphics{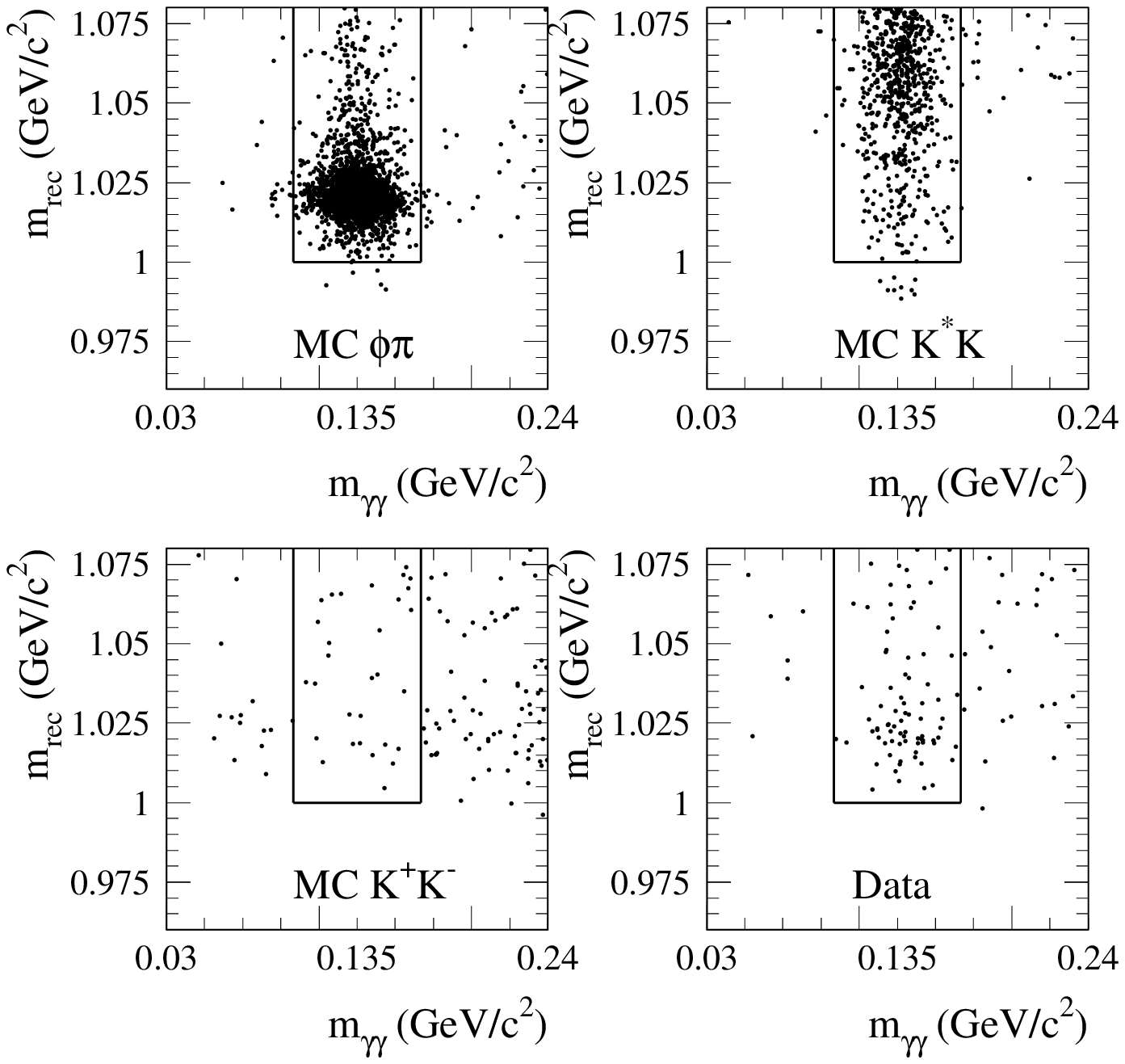}}
\caption{\label{mrecpi0vsm12k}
The two-dimensional $m_{\rm rec}$ versus $m_{\gamma\gamma}$ distribution
for selected data and simulated events of the
processes $e^+ e^- \to \phi \pi^0 \to K^+ K^- \pi^0 $,
$e^+ e^- \to K^\ast K \to K^+ K^- \pi^0$, $e^+ e^- \to K^+ K^- (\gamma)$.
The lines indicate the region of invariant masses
 ($1.00 < m_{\rm rec}< 1.08$ GeV/$c^2$, $0.1 <  m_{\gamma\gamma} < 0.17$
  GeV/$c^2$) used in the $e^+e^- \to \phi \pi^0$ analysis.
}
\end{figure*}

\begin{figure}
\centering
\resizebox{0.45\textwidth}{!}{\includegraphics{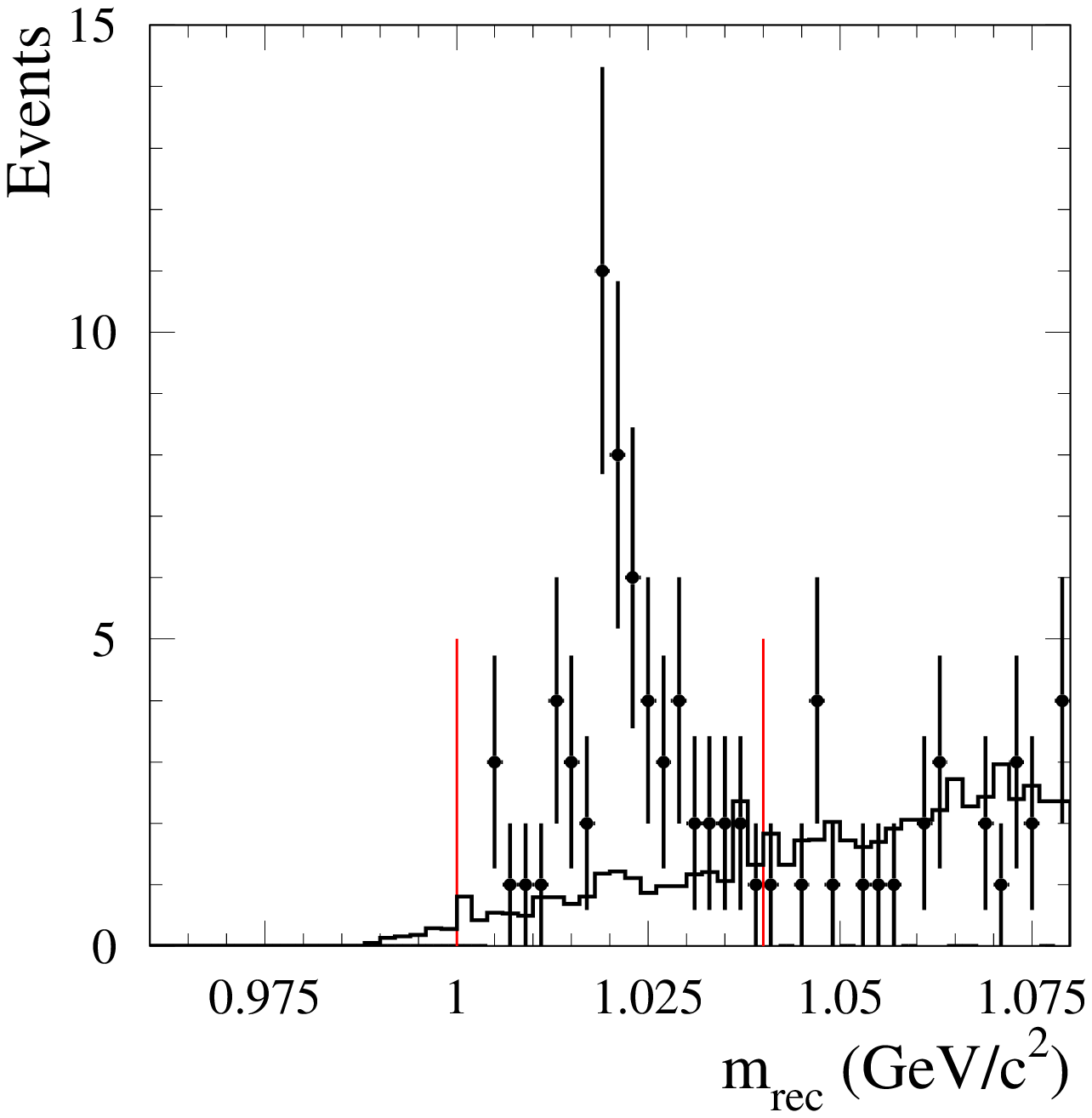}}
\caption{
The $m_{rec}$ distributions for data events (points with error bars).
The histogram represents a sum of the simulated distributions for
$e^+e^-\to  K^\ast K \to K^+ K^-\pi^0 $ events and events of the 
background processes~(\ref{bkg_proc}). The vertical lines indicate
the region $1.00<m_{\rm rec}<1.04$ GeV/$c^2$ used for measurement of the
$e^+ e^- \to \phi \pi^0 \to K^+ K^- \pi^0$ cross section.
\label{mrecpi0vsm12k1}}
\end{figure}

Figure~\ref{mrecpi0vsm12k} shows the two-dimensional distributions of 
$m_{\rm rec}$ versus $m_{\gamma\gamma}$ for data events, simulated 
$e^+ e^- \to \phi \pi^0 \to K^+ K^- \pi^0 $ events, and simulated
events of the main background processes,
$e^+ e^- \to K^\ast K \to K^+ K^- \pi^0$ and $e^+ e^- \to K^+ K^-(\gamma)$.
Figure~\ref{mrecpi0vsm12k1} shows the $m_{\rm rec}$ spectrum for data events
with $0.1<m_{\gamma\gamma}<0.17$~GeV/$c^2$, in which the 
$\phi (1020)$ peak is clearly seen. The expected distribution for
background events is also presented. 
It is seen that the simulation reproduces well both the total number of
background events and the shape of the background distribution.

We define the signal ($1.00<m_{\rm rec}<1.04$ GeV/$c^2$) and sideband 
($1.04<m_{\rm rec}<1.08$ GeV/$c^2$) mass regions and determine the number of
$e^+ e^- \to \phi \pi^0 \to K^+ K^- \pi^0$ events using the equation
\begin{equation}
N = \frac{N_1 - k_b*N_2}{1-k_s*k_b},
\end{equation}
where $N_1$ and $N_2$ are the numbers of data events in the signal and
sideband regions, respectively, $k_b$ is the $N_1/N_2$ ratio
for background events, and $k_s$ is
the $N_2/N_1$ ratio for signal events. The coefficients $k_b$ and $k_s$
are determined from simulation.

The detection efficiency for $e^+e^- \to \phi\pi^0 \to K^+K^-\pi^0$
events obtained using MC simulation grows
from 1\% at $\sqrt{s}=1.4$~GeV to 8\% at $\sqrt{s}=1.8$ GeV,
and then decreases to 6\% at $\sqrt{s}=2$ GeV.

\begin{figure}
\centering
\resizebox{0.45\textwidth}{!}{\includegraphics{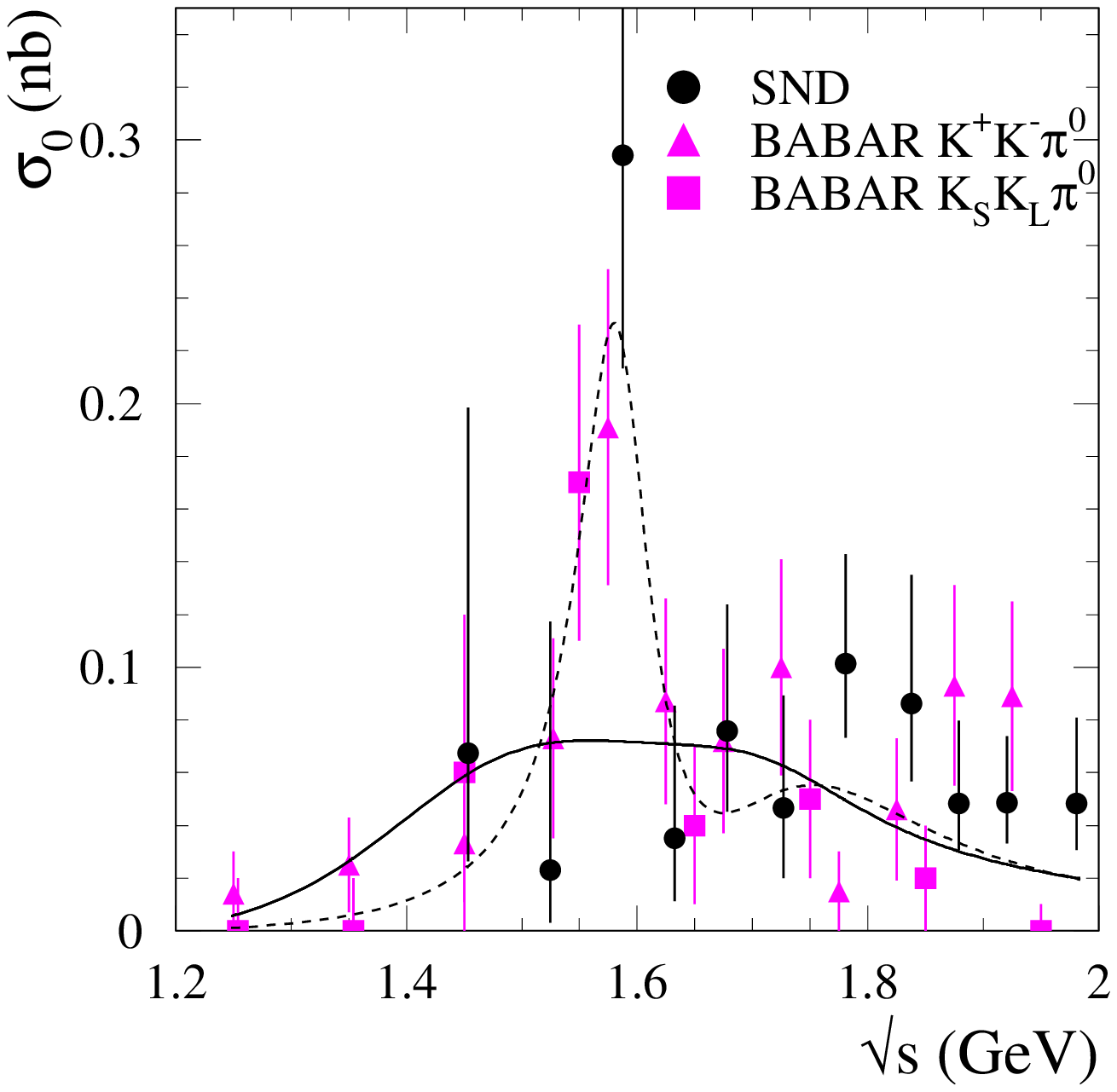}}
\caption{
The cross section for the process $e^+e^- \to \phi \pi^0\to K^+K^-\pi^0$
obtained in this experiment in comparison with the two BABAR 
measurements~\cite{bib:babar1,bib:babar2}.
The solid and dashed curves represent the results of the fit in Models I
and II, respectively.
\label{crphipi0_700}}
\end{figure}

To calculate the radiative corrections and experimental values of
the Born cross section, we perform simultaneous fit to the SND data and 
the data from the two BABAR measurements~\cite{bib:babar1,bib:babar2}.
The Born cross section is described by the coherent sum of the contributions
of the $\rho (1450)$ and $\rho(1700)$ resonances (Model I).
In this model, the masses and widths of the resonances are fixed at the
PDG values~\cite{pdg}, while the cross sections at the resonance maxima and
the relative phase between the resonance amplitudes are free fit parameters. 
The obtained values of the Born cross section for the process
$e^+e^- \to \phi \pi^0\to K^+K^-\pi^0$ are listed in Table~\ref{table2}
and are shown in Fig.~\ref{crphipi0_700} together with the BABAR data
and the fitted curve. It is seen that all three measurements are in good 
agreement below 1.75 GeV. In the range 1.75--2~GeV
the nonstatistical spread of the measurements is observed.
The fitted curve agrees with the data everywhere except in the narrow
region near $\sqrt{s}=1.58$~GeV, where excess over the curve is observed in
all three measurements. The overall fit quality is unsatisfactory
($\chi^2/{\rm ndf}=50/28$).
\begin{table*}
\centering
\caption{\label{table2}
The measured cross section for the process $e^+ e^-\to \phi \pi^0\to
K^+K^-\pi^0$ as a function of the c.m. energy $\sqrt{s}$. The quoted 
errors are statistical.
}
\begin{tabular}{cccccc}
\hline
$\sqrt{s}$ GeV) &$\sigma$ (nb) & $\sqrt{s}$ (GeV)&$\sigma$ (nb)&
$\sqrt{s}$(GeV) & $\sigma$ (nb)\\
\hline
 $1.40 - 1.50 $ & $ 0.033^{+0.064}_{- 0.020} $ &$1.65 - 1.70 $ & $ 0.037^{+0.024}_{-0.015} $ &$1.85 - 1.90 $ & $0.024^{+0.015}_{-0.009} $ \\
 $1.50 - 1.55 $ & $ 0.011^{+0.046}_{- 0.010} $ &$1.70 - 1.75 $ & $ 0.023^{+0.021}_{-0.013} $ &$1.90 - 1.95 $ & $0.024^{+0.012}_{-0.008} $ \\
 $1.55 - 1.60 $ & $ 0.145^{+0.054}_{- 0.040} $ &$1.75 - 1.80 $ & $ 0.050^{+0.020}_{-0.014} $ &$1.95 - 2.10 $ & $0.024^{+0.016}_{-0.009} $  \\
 $1.60 - 1.65 $ & $ 0.017^{+0.025}_{- 0.012} $ &$1.80 - 1.85 $ & $ 0.042^{+0.024}_{-0.014} $ & & \\		     
\hline
\end{tabular}
\end{table*}

A better description of the data is obtained with the two resonance model,
in which the mass and width of the first resonance are fixed at the PDG 
values for the $\rho(1700)$, and the parameters of the second resonance
are free (Model II). The fit in this model yields $\chi^2/{\rm ndf} =
38/26$ ($ P(\chi^2) = 6\%$), and the following parameters of the second 
resonance: $M=1585 \pm 15$~MeV and $\Gamma=75 \pm 30$~MeV.
The fitted curve for Model II is also shown in Fig.~\ref{crphipi0_700}.
It should be noted that there is no a vector resonance with such parameters
in the PDG table~\cite{pdg}. Formally, its significance calculated from
the difference of the $\chi^2$ values for Models I and II is about $3\sigma$.

The difference in the radiation corrections calculated with Models I
and II is used to estimate the model uncertainty on the Born cross
section. It is 14\% for the interval 1.6--1.65~GeV, 8\% for the interval
1.65--1.7~GeV, and does not exceed 6\% for the remaining points.
The systematic uncertainty on the cross section is similar to that
for the $e^+e^-\to K^+K^-\pi^0$ cross section and does not exceed 10\%.

The intermediate state $K^\ast K$  gives nonzero contribution to
the signal region $1.00<m_{\rm rec}<1.04$ GeV. This leads to 
interference between the $\phi\pi^0$ and $K^\ast K$ amplitudes,
which may contribute to the measured $e^+ e^-\to \phi \pi^0 \to K^+K^-\pi^0$
cross section. Using a model with a coherent sum of the $\phi\pi^0$ and
$K^\ast K$ amplitudes we vary the phase difference between them and study how
the interference modifies the $m_{\rm rec}$ spectrum. It is found that using
the procedure of the $\phi\pi^0$ signal extraction described above 
we actually measure a sum of the $e^+ e^-\to \phi \pi^0 \to K^+K^-\pi^0$
cross section and the interference term integrated over the $m_{\rm rec}$ 
signal region with an uncertainty of 30\%.

To understand how large the effect of the interference is,
we fit the $e^+ e^-\to \phi \pi^0 \to K^+K^-\pi^0$ cross section measured
in this work and by BABAR~\cite{bib:babar1} with the following model:
\begin{eqnarray}
\sigma_{\rm meas}(s)&=&\sigma_{\phi\pi^0}(s)+
2\sqrt{\sigma_{\phi\pi^0}(s)\sigma_{K^\ast K}(s)}\nonumber\\
&\times&
\big ( O_{\rm Re}(s)\cos{(\psi+\psi_{K^\ast K}}-\psi_{\phi\pi^0})\nonumber\\
&+&
O_{\rm Im}(s)\sin{(\psi+\psi_{K^\ast K}-\psi_{\phi\pi^0})}\big ),\label{eq8}
\end{eqnarray}
where $\sigma_{\phi\pi^0}(s)$ and $\sigma_{K^\ast K}(s)$ are the cross sections
corresponding to the squared moduli of the $\phi\pi^0$ and $K^\ast K$
amplitudes, respectively, $\psi_{\phi\pi^0}(s)$ and $\psi_{K^\ast K}(s)$ are
the arguments of these amplitudes, $O_{\rm Re}(s)$ and $O_{\rm Im}(s)$
are the real and imaginary parts of the specially normalized overlap integral
between the $\phi\pi^0$ and $K^\ast K$ amplitudes, and $\psi$ is the relative
phase between them. 
\begin{figure}
\centering
\resizebox{0.45\textwidth}{!}{\includegraphics{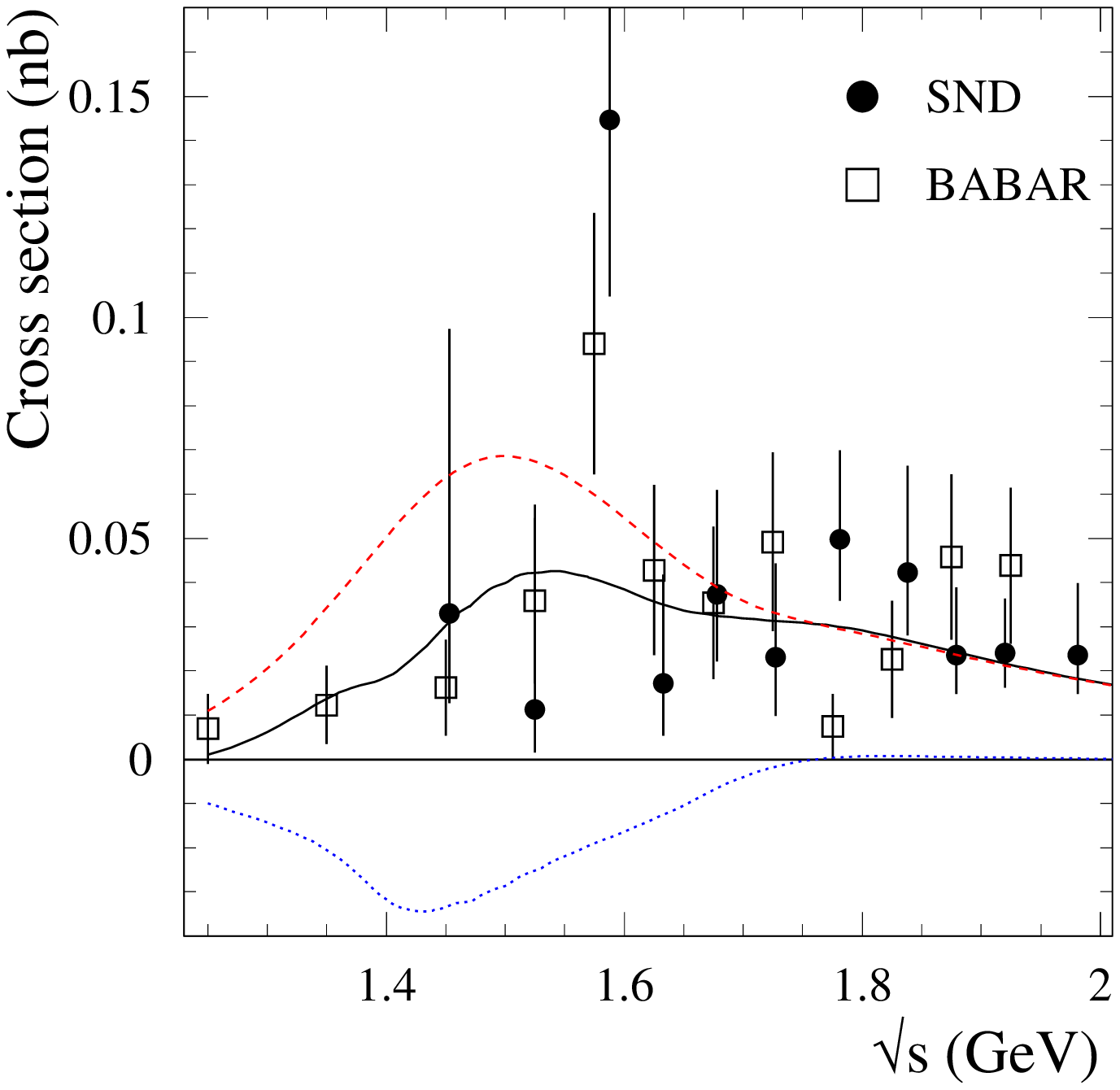}}
\caption{
The cross section of the process $e^+e^- \to \phi \pi^0\to K^+K^-\pi^0$
obtained in this work and in the BABAR experiment~\cite{bib:babar1}.
The solid curve is the result of the fit to the cross section data 
with Eq.~(\ref{eq8}). The dashed and dotted curves represent 
the $\sigma_{\phi\pi^0}$ term and the interference terms of Eq.~(\ref{eq8}),
respectively. 
\label{interf}}
\end{figure}
The functions $\sigma_{K^\ast K}(s)$ and $\psi_{K^\ast
K}(s)$ are determined from the fit to the $e^+ e^-\to K^+K^-\pi^0$ cross
section as described in Sec.~\ref{kstkfit}. The $\phi\pi^0$ amplitude is
parametrized using Model I introduced above. An additional fit parameter is
the phase $\psi$. The result of the fit is shown in Fig.~\ref{interf}.
The energy dependence of the fitted $\sigma_{\phi\pi^0}$ and interference
terms are also shown. 

It is seen that the interference with the $K^\ast K$ amplitude gives 
sizable contribution to the measured $e^+e^- \to \phi \pi^0 \to K^+K^-\pi^0$
cross section listed in Table~\ref{table2}. Below 1.7 GeV the measured cross 
section cannot be directly associated with the $e^+e^- \to \phi \pi^0$ cross
section. 

The fitted curve in the model with interference does not differ
significantly from the curve obtained in the model without interference 
(Model I in Fig.~\ref{crphipi0_700}). Both  models cannot reproduce the 
narrow structure near 1.6 GeV seen in the SND and two BABAR measurements.

The total $e^+e^- \to K^+K^-\pi^0$ cross section can be calculated by summing
the cross sections listed in Tables~\ref{table1} and \ref{table2}. The resulting
cross section accounts for the interference between the $K^\ast K$ and
$\phi\pi^0$ intermediate states. 

\section{Summary}
In this paper the process $e^+e^-\to K^+K^-\pi^0$ has been studied
in the c.m. energy range from 1.28 to 2 GeV.
We have analyzed the data with an integrated luminosity 26.4~pb$^{-1}$
accumulated in the experiment with the SND detector at the 
VEPP-2000 $ e^+ e^-$ collider in 2011-2012.
It has been shown  that the process $e^+e^-\to K^+K^-\pi^0$ in the 
energy range under study proceeds predominantly 
through the $K^{\ast}(892)^{\pm} K^{\mp}$ intermediate state.
The signal from the intermediate state $\phi\pi^0$ has been also
observed. The cross sections for the process $e^+e^-\to K^+K^-\pi^0$ (without
$\phi \pi^0$) and $e^+e^- \to \phi \pi^0\to K^+K^-\pi^0$ have been measured
separately. They agree well with the previous
measurements in the BABAR experiment and have comparable accuracy.

For the process $e^+e^- \to \phi \pi^0\to K^+K^-\pi^0$ we have studied the
effect of the interference between the $\phi\pi^0$ and $K^\ast K$ amplitudes.
It has been found that the interference gives sizable contribution
(up to 100\%) to the measured $e^+e^- \to \phi \pi^0\to K^+K^-\pi^0$ cross 
section below 1.7 GeV. In this region we actually measure the sum of the
$\phi\pi^0$ cross section and the interference term with the model uncertainty
of 30\%. Within this uncertainty, the total
$e^+e^- \to K^+K^-\pi^0$ cross section calculated as a sum of the two measured
cross sections accounts correctly for the interference between the 
$\phi\pi^0$ and $K^\ast K$ amplitudes.

In the narrow region near $\sqrt{s}=1.58$~GeV all three existing measurements
of the $e^+e^- \to \phi \pi^0$ cross section, performed by SND (this work) and
BABAR~\cite{bib:babar1,bib:babar2}), show excess over the model including
known vector resonances. This excess can be interpreted as a contribution of
the resonance with $M=1585 \pm 15$~MeV and $\Gamma=75 \pm 30$~MeV. Its 
significance is estimated to be about $3\sigma$.

\acknowledgement{
The work was performed using the unique scientific facility 
``Complex VEPP-4 -- VEPP-2000''.}

%\newpage

\end{document}